# Convolutional neural networks enable high-fidelity prediction of path-dependent diffusion barrier spectra in multi-principal element alloys

Zhao Fan[1,*], Bin Xing[2], Penghui Cao[1,*]
[1]Department of Mechanical and Aerospace Engineering, University of California, Irvine, Irvine, California 92697, United States
[2]Department of Materials Science and Engineering, University of California Irvine, Irvine, CA 92697, USA
[*] Emails: zfan2016@gmail.com;  caoph@uci.edu

## Abstract

The emergent multi-principal element alloys (MPEAs) provide a vast compositional space to search for novel materials for technological advances. However, how to efficiently identify optimal compositions from such a large design space for targeted properties is a grand challenge in material science. Here we developed a convolutional neural network (CNN) model that can accurately and efficiently predict path-dependent vacancy migration energy barriers, which are critical to diffusion behaviors and many high-temperature properties, of MPEAs at any compositions and with different chemical short-range orders within a given alloy system. The success of the CNN model makes it promising for developing a database of diffusion barriers for different MPEA systems, which would accelerate alloy screening for the discovery of new compositions with desirable properties. Besides, the length scale of local configurations relevant to migration energy barriers is uncovered, and the implications of this success to other aspects of materials science are discussed.

# Introduction

Technological advances in civilization usually are driven by the emergency of novel materials. Traditional alloying approaches based on the paradigm of "one-dominant-element" have been utilized to impart desired properties (e.g., high strength and high ductility) to materials for millennia[1]. Recently, a new alloy design concept[1-3] that combines multiple principal elements in high concentration has emerged. It was conceived in 2004 that mixing multiple elements with equal concentrations could form a simple solid solution[4,5]. This brings about a vast unexplored territory to search for new materials with target properties and thus triggered the surge in research activity. These multi-principal element alloys (MPEAs), commonly known as HEAs (high-entropy alloys) or CCAs (complex concentrated alloys), have been shown to possess exceptional mechanical and functional properties, including high strength combined with large ductility[6,7], high fracture toughness[8,9], thermoelectric properties[10,11], enhanced radiation tolerance[12,13] and extraordinary Elinvar effect[14]. Many of these properties and behaviors, particularly at high temperatures, such as weak temperature dependence[15] and high creep tolerance[2], are mainly governed by diffusion kinetics.

Understanding diffusion kinetics[16] is of importance to evaluate the phase stability and high-temperature deformation characteristics of HEAs, which are critical to assess the performance of materials[17]. Therefore, diffusion and vacancy kinetics in HEAs are currently a subject of intensive research[17-33]. Vacancy migration energy barrier (VMEB), one of the key parameters related to diffusion kinetics, can be calculated from atomistic models using transition state calculations such as climbing image nudged elastic band (NEB)[34]. For a single crystal of pure metals, its VMEB is a single value. But for HEAs, their VMEBs form wide spectra (distributions) due to the diverse local chemical environments surrounding vacancies[28-32]. Even for a given HEA with a fixed composition, its VMEB spectrum and thus diffusion kinetics can vary[32] through tuning the chemical short-range order (CSRO)[35,36]. Therefore, it is desirable to build a big database capturing all VMEB spectra for HEAs at different compositions and/or with various degrees of CSRO. This would be valuable for material scientists to quickly identify ideal compositions from the vast compositional space and optimize processing conditions to achieve diffusion-related properties for various applications. However, developing such databases directly using standard methods such as NEB method[34] is a cost-prohibitive task because of the hyper-compositional space and locally diverse chemical environments in HEAs.

Recently, convolutional neural networks (CNNs) have brought about revolutionary breakthroughs in the field of computer vision and pattern recognition[37-40]. Inspired by that, here we aim to develop a CNN model which can accurately and efficiently predict VMEB spectra for all alloys at different compositons and with various degrees of CSRO within a given multiple-element alloy system, solely based upon local chemical environments. In this deep learning (DL) framework, local chemical environments around vacancies will be described using a new structure representation—spatial density maps (SDMs)[41], which are equivalent to three-dimensional (3D) images and thus can be interpreted directly by CNNs. In addition to the rotationally non-invariance of SDMs, which can capture the path-dependence of VMEB in HEAs due to the asymmetry of local chemical environments, the completeness[42] of SDMs as well as the state-of-the-art learning capability of CNNs will ensure this DL framework is able to achieve impressive accuracy when predicting VMEB in HEAs. Our DL framework will thus make it possible to build spectral vacancy migration databases for different multiple-element alloy systems with affordable computational cost. These databases will be a general and broadly applicable toolbox for alloy design and processing optimization relevant to diffusion-governed behaviors.

## Results

**Atomic structure representation and CNN model architecture.** Here, a body-centered cubic (BCC) refractory ternary alloy system, Ta-Nb-Mo, modeled with a machine learning potential[43] is chosen to demonstrate the feasibility of our DL framework. Fig. 1a shows a global configuration of an equal-atomic random TaNbMo atomistic model which is projected on the *xy* plane. Different from almost all previous machine learning tasks of predicting local properties of materials, such as predicting grain boundary segregation energy in polycrystals[44] or flexibility volume in metallic glasses[45], where the target values of interest are rotation-invariant, our target values, i.e., VMEBs, are *path-dependent*, as there are eight first-nearest neighboring atoms around a vacancy in a BCC alloy and the barriers of their migrations to the vacancy should be different due to the asymmetry of local chemical environment in HEAs. So, it is improper to apply the structure representations widely used in previous works, e.g., symmetry functions[46] and smooth overlap of atomic positions (SOAP)[47], to predict path-dependent migration energy barriers in our current work, because these structure representations are rotation-invariant.

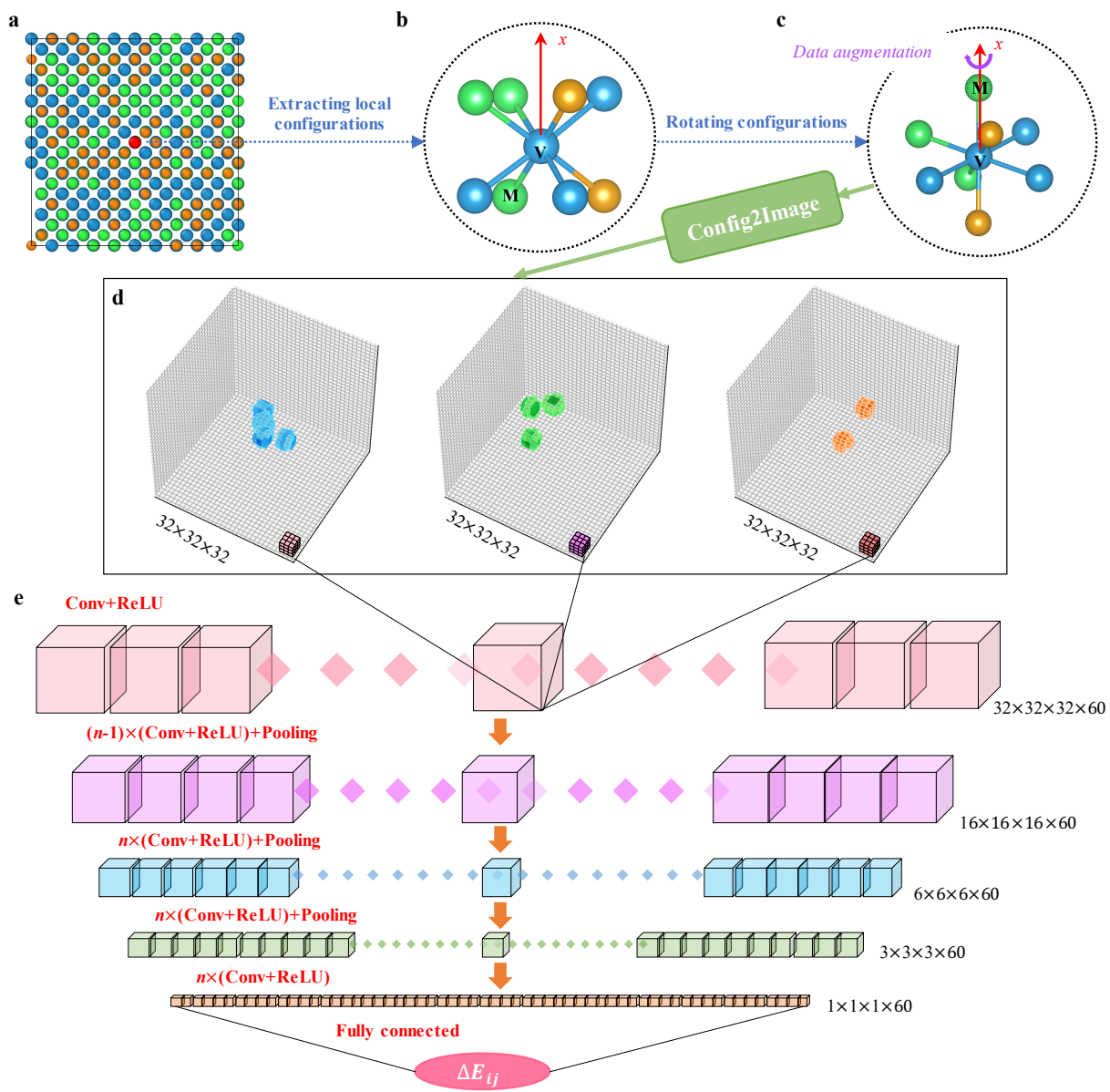

**Fig. 1 Atomic structure representation and CNN model architecture. a** The two-dimensional projection view of a random solid solution of equal-atomic TaNbMo containing 2,000 atoms. Blue, green and orange spheres represent Ta, Nb and Mo species, respectively. **b** A local configuration around the vacancy in **a** (red dot). For clarity, only eight first-nearest neighbors are shown. There are eight migration energy barriers associated with the vacancy "V", $\Delta E_{VM}$, where "M" indicates the first-nearest neighbor which would migrate to the vacancy. **c** To predict path-dependent barriers, the local configuration is rotated such that the migration vector of interest is aligned with a given direction, here we choose $x$ axis. For data augmentation, the local configurations in training datasets is rotated around $x$ axis by a random angel in the range of $[0, 2\pi)$ before converted into spatial density maps (SDMs). **d** shows the three channels corresponding to Ta, Nb and Mo

species, respectively, of the SDM for the local configuration shown in **c**. Only voxels with intensity greater than 0.2 are shown for clarity. **e** An illustration of the architecture of the convolutional neural network (CNN) used in this work, explicitly delineating the output features after the first and last convolutional (conv.) layers and three max-pooling layers in the middle. The first convolutional layer connects locally with SDMs through 60 filters, one of which is shown on the corners of the three channels in **d**. And the last convolutional layer is followed directly by the output layer which is a single neuron and thus the output is a single value, i.e., the migration barrier.

In this work, we use a rotationally non-invariant local structure representation—spatial density map (SDM)[41]—to describe the local chemical environments enclosing vacancies in alloys. The SDM centered on each vacancy site is defined as

$$\Xi_i(z, y, x, \beta) = \sum \exp\left(-\frac{(r_{ij,x}-x)^2+(r_{ij,y}-y)^2+(r_{ij,z}-z)^2}{2\Delta^2}\right), \quad (1)$$

where the summation is performed over all atoms satisfying these conditions: species$(j) \in \beta$ and $|r_{ij}| < r_c$. $\boldsymbol{r}_{ij}$ is the vector connecting vacancy $i$ with surrounding atom $j$ of species $\beta$ within a cutoff $r_c$, and $r_{ij,x}$, $r_{ij,y}$ and $r_{ij,z}$ are the components along x, y and z dimensions of $\boldsymbol{r}_{ij}$, respectively. Here $\beta \in \{0, 1, 2\}$, where 0, 1 and 2 represent Ta, Nb and Mo atoms, respectively, and $x, y, z \in [-l_c + 0.5\Delta, l_c - 0.5\Delta]$ with a constant length increment of $\Delta$. $l_c$ decides the size of SDMs or images and should be equal to or slightly larger than $r_c$, which is the radius of spherical local configurations. In the previous work[41], local configurations in cubic boxes were input into a machine and $r_c$ and $l_c$ are actually the same parameter $r_c$. In the current work, we will augment training data through rotating local configurations around a given axis, which will be discussed later. Thus, here we are considering spherical local configurations and the conditions defining atoms involved in the summation of equation (1) are different from those in the previous work. To minimize the variation of output values of DL models due to rotation of local configurations, it is better to set the half edge length of cubic SDMs slightly larger than the radius of local configurations, i.e., $l_c$ is slightly larger than $r_c$.

Using an SDM to represent a local atomic configuration can be viewed as mapping the local configuration to a 3D grid for different species. This results in a multi-dimensional numerical array which is equivalent to a 3D image containing $(2 l_c/\Delta)^3$ voxels. And each voxel has channels equal to the number of components in samples. Thus, the SDM is applicable to HEA systems containing any number of constituent elements through adjusting the number of channels. The larger the $r_c$ is, the more surrounding particles would be included into an SDM. And when $\Delta$ is small enough, any tiny variation of particle positions in the local configuration would lead to a corresponding variation in its SDM. Therefore, SDMs can represent the full gene of both chemical and structural information of local configurations when $r_c$ is large enough and $\Delta$ is small enough. Although a larger $r_c$ as well as a smaller $\Delta$ can ensure the completeness of SDMs for the representation on local atomic environments, which is a critical factor for the success of machine learning models[42], a too large $r_c$ or a too small $\Delta$ also means much larger size of each image. This would require more computational resource and memory to generate, store and utilize these images. In the next section, we will discuss how to choose appropriate values for $r_c$ and $\Delta$. Note that each SDM or image represents the *local* chemical environment centered on the site of interest rather than a global sample. And the SDM is naturally permutation-invariant because of the summation operation in equation (1).

Fig. 1b shows a local configuration around the vacancy (red colored dot in Fig. 1a). For clear illustration, only eight first-nearest neighbors are shown in this plot. In our actual local environment representation, we used a larger $r_c$ to include sufficient neighbors for each local configuration. The target value of our DL task is the vacancy migration energy barrier $\Delta E_{ij}$, the true value of which was calculated using the NEB method[34] (see *Methods* for more details). $\Delta E_{ij}$ measures the migration energy barrier of the first-nearest neighbor $j$ (e.g., the atom marked with "M") to the central vacant site $i$. Because all the eight first-nearest neighbors are possible to migrate to this vacancy, each vacancy has eight migration energy barriers (eight migration pathways). As illustrated in Fig. 1c, before converting the local configuration associated with each $\Delta E_{ij}$ into an SDM, the local configuration was rotated such that the vector connecting the vacancy "V" with moving atom "M" is parallel to *x* axis. Therefore, when considering the barrier of moving each of the eight nearest atoms to the same vacant site, we just need to conduct appropriate rotation based on the convention described above before converting the local configuration into an SDM and then a DL model will naturally give us all the path-dependent energy barriers for the vacancy $i$.

Obviously, arbitrarily rotating a local configuration around *x* axis will not change its associated

$\Delta E_{ij}$. We augmented our training datasets based on this physic intuition. Specifically, all local configurations in the training dataset were rotated around *x* axis by a random angle over the range of [0, 2π) before finally converted into SDMs. Fig. 1d shows the three channels corresponding to Ta, Nb and Mo species, respectively, of the SDM for the local configuration displayed in Fig. 1c.

We then feed each SDM into a CNN model[40] to predict $\Delta E_{ij}$, since SDMs are image-like objects and can be used directly as input into CNN models. A CNN model mainly consists of convolutional and pooling layers. CNNs are very powerful as their learning capability can be effectively elevated by deepening the networks, i.e., stacking more convolutional layers (including residual connections[39] may be necessary when a network is very deep). As shown in Fig. 1e, each of our CNN models in the present work contains 4×*n* 3D convolutional layers (various values of *n* will be explored) and three 3D max-pooling layers. In each convolutional layer, 60 filters with a small receptive field of $3 \times 3 \times 3$ were used. Batch normalization[48] was adopted right after each convolution and before activation with the rectification (ReLU) nonlinearity[37]. A 3D max-pooling layer is periodically inserted in between the 4×*n* successive convolutional layers. Max-pooling is performed over a 2×2×2 voxel window, with a stride of 2. And the last convolutional layer is directly followed by the output layer, which is a single neuron without any activation as we are performing regression tasks.

In the regression tasks, we use the mean absolute error (MAE), i.e., the averaged absolute value of the difference between the true and predicted target values ($\Delta E_{ij}$) over a given dataset, to measure the predictive performance of CNN models. During the training stage, the training dataset will be fed iteratively into the CNN model to optimize its trainable parameters until the lowest MAE on the validation dataset is observed. Once an optimized CNN model is obtained after sufficient training, it can quickly and accurately predict the migration energy barriers for alloys at arbitrary compositions within the alloy system upon which the CNN model is trained, with negligible computational resource. The sole input to the CNN model is local configurations, which can be easily extracted from atomistic models. The computation-expensive NEB calculation will not be required any more.

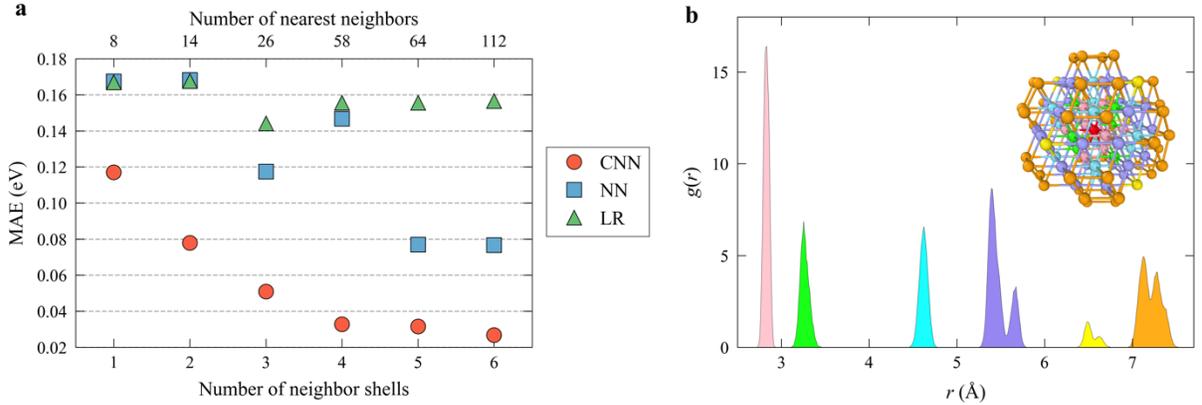

**Fig. 2 The length scale of local configuration relevant to migration energy barrier. a** The validation mean absolute error (MAE) as a function of number of neighboring shells. Three machine learning methods including CNN, neural network (NN) and linear regression (LR) are trained and validated on an equimolar TaNbMo alloy. **b** The radial density function ($g(r)$) of the equimolar TaNbMo alloy. The peaks corresponding to different neighboring shells are filled with different colors. The inset shows a local configuration around the central vacancy (red dot) containing six nearest neighbor shells (112 atoms). The atoms in different neighboring shells are coded by different colors.

**The length scale of local configurations relevant to migration energy barriers.** Before training a CNN model for the entire compositional space of the ternary Ta-Nb-Mo alloy system, we first reveal how many neighboring atoms around a vacancy can impact its migration energy barrier and should be fed into a machine learning model. For this purpose, a small training/validation dataset was constructed at a single composition of the equal-atomic TaNbMo alloy. Specifically, a random solid solution and an alloy with CSRO were created. The latter was generated based on the first using the hybrid Monte Carlo (MC)/molecular dynamics (MD) simulation (see *Methods*). Each of the two samples contains 4,800 atoms and for each sample, 800 atoms were selected randomly and removed individually to create a vacancy and then the migration barrier of each of the eight first-nearest neighbors to the vacancy was calculated using the NEB method (the removed atom will move back to the vacancy once the calculation of barriers associated with the vacancy is done). Thus, there are totally 12,800 instances (2 samples × 8 paths × 800 vacancies). These data were split into the training and validation datasets as a ratio of 4:1. When converting local configurations into SDMs, six different values of $r_c$ were tried to check its influence on the predictive performance of CNN models. For BCC crystals, the neighboring shells are well separated, see Fig. 2b, and thus, $r_c$ can be any value at which the radial density function $g(r) = 0$ as long as it is between the $n$th and ($n$+1)th peak locations of $g(r)$, when considering the first $n$ nearest-neighboring shells. For example, when we included six nearest-neighboring shells, $r_c$ = 7.6 Å ($l_c$ = 8.0 Å) was used. To avoid too large image size, $\Delta$ = 0.5 Å was used when $l_c$ = 8.0 Å. With such values of $l_c$ and $\Delta$, each SDM contains $32^3$ voxels. As will be demonstrated in Fig. 2a, $\Delta$ = 0.5 Å is sufficiently small to achieve low MAE. And for smaller $l_c$, $\Delta$ = 0.5 or 0.4 Å was used.

For each of the six training/validation datasets, we separately trained a CNN model. As seen from Fig. 2a, the validation MAE decreases when more neighboring shells are included and the MAE almost converged when six nearest-neighboring shells were considered. This implies that the VMEB is mainly governed by the chemical environment within the first six nearest-neighboring shells, which include 112 nearest-neighboring atoms.

To show the advantage of our CNN models in achieving high-fidelity prediction and revealing reliable physical insight, we designed a simple descriptor to represent local chemical order in HEAs and then used two conventional machine learning algorithms, i.e., a linear regression (LR) and a fully-connected neural network (NN), to interpret the datasets constructed uisng the new simple descriptor. Specifically, after using various $r_c$ to define local configurations and rotating them to the orientation shown in Fig. 1c, we use a three-element hot vector to describe the species of each lattice site and stack these hot vectors as a given sequence into a matrix for each local configuration. Obviously, these matrixes can fully represent the chemical order information of local configurations despite the loss of topological information, which should be trivial to migration energy barriers in HEAs. They will be reshaped into vectors before fed into a LR or NN model.

Fig. 2a also shows the validation MAE of the two machine learning frameworks. For the NN model, its MAE is much higher than that achieved by the CNN model on the same dataset. Besides, although it also suggests the overall trend that including more neighboring shells can further lower the validation MAE, there is an abnormality when including the first four neighboring shells (10 independent NN models were trained at this point but resulted in almost the same MAE). This comparison highlights the advantage of using state-of-the-art machine learning algorithms to achieve higher predictive power (or lower MAE for the current case).

Except the much poor predictive performance, the LR model achieved its lowest MAE when including the first three nearest-neighboring shells, which should be ascribed to its weak learning capability. This comparison highlights the importance of using advanced machine learning frameworks when extracting physical insight from machine learning results. Otherwise, some misleading conclusions might be reached. For example, one might conclude that the chemical order beyond the first three nearest shells is irrelevant to migration energy barriers in HEAs if only the LR model was utilized for the current case.

**High-fidelity prediction on the entire compositional space**. In order to train a CNN model robust for all alloys at different compositions and/or with different degrees of CSRO within the ternary Ta-Nb-Mo alloy system, we created 46 samples with different compositions, which uniformly occupy the ternary Ta-Nb-Mo compositional space, see supplementary Fig. 1, to construct both training and validation datasets. All these samples are random solid solutions, each containing 2,000 atoms. The eight barriers for each of the 2,000 possible vacant sites in each sample were calculated using the NEB method[34]. So there are 16,000 barriers at each composition. 218 barriers were selected at random from each composition to construct the validation dataset (totally 10,028 instances) and the remaining barriers at all compositions were used to construct a big training dataset, which totally contains 725,972 instances. Note that although the samples used to construct training/validation datasets are random solid solutions, the possible CSROs in the concentrated alloys should be similar as those in the random dilute alloys and the CNN model trained on this group of datasets is thus expected to be valid for the concentrated alloys with various degrees of CSRO, which will be verified later.

We then trained several CNN models with different depths on the same training dataset. Supplementary Fig. 2 shows the both training and validation MAE as a function of number of convolutional layers contained in CNN models. When only four convolutional layers are used, both training and validation MAE are relatively high, suggesting that a deeper CNN model is needed. With inceasing the number of convolutional layers, both the training and validation MAE decrease effectively. When the number of convolutional layers increases to twelve, the lowest validation MAE of 0.0137 eV is observed and both training and validation MAEs are equal to each other. It is observed that the training MAE is slightly lower than the validation MAE when further increasing the number of convolutional layers to sixteen. These results suggest that for the current training dataset, a CNN model containing twelve convolutional layers is sufficient and further deepening the CNN model would result in overfitting. We believe that an even lower validation MAE can be achieved through constructing larger training datasets and using deeper CNN models. The main purpose of this work is to demonstrate the feasibility of our DL framework.

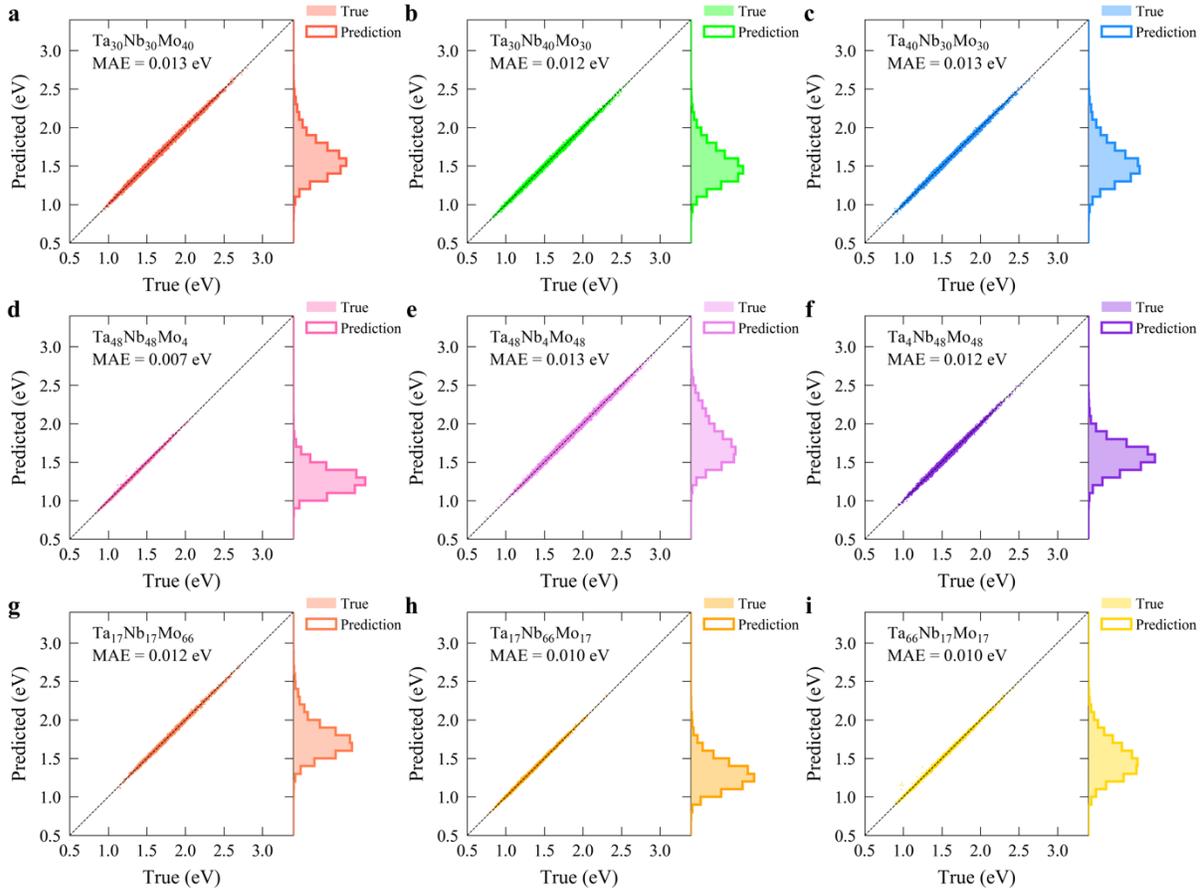

**Fig. 3 Test of the high-fidelity CNN model across the compositional space.** The predictive power—mean absolute error (MAE)—of the CNN model trained on the big dataset for vacancy migration energy barriers in nine alloys at different compositions, which are never used in either training or validation dataset. The specific MAE for each composition is denoted on each panel. Each of the first six samples (**a-f**) contains 2,000 atoms (16,000 barriers), while each of the last three samples (**g-i**) contains 686 atoms (5,488 barriers).

We termed the optimized CNN model containing twelve convolutional layers high-fidelity CNN model and then tested it on alloys at nine different compositions which are never seen at the training and validation stage (see supplementary Fig. 1). As exhibited in Fig. 3, the test MAE on the nine compositions are comparable to or even lower than the validation MAE and there is almost no difference between the true and predicted distributions of $\Delta E_{ij}$. This confirms that our CNN model trained using sparse composition data (46 samples) is indeed able to capture the entire compositional space of the Ta-Nb-Mo system. To showcase our CNN model is also robust to samples with different size across the entire ternary compositional space, we especially used small samples each only containing 686 atoms for the last three compositions in Fig. 3g-i. We also compared the true $\Delta E_{ij}$ distribution of a small equal-atomic TaNbMo alloy containing 2,000 atoms and the predicted $\Delta E_{ij}$ distribution of a very large equal-atomic TaNbMo alloy containing 1,024,000 atoms. The predicted distribution of the very large sample is consistent to the true distribution of the small sample, see supplementary Fig. 3.

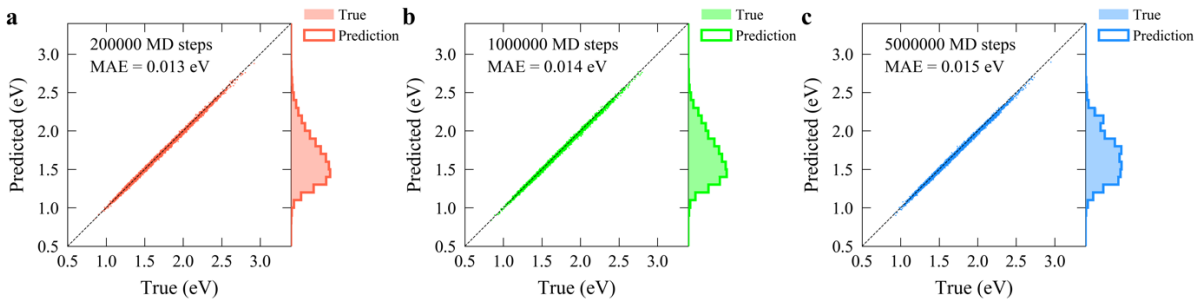

**Fig. 4 Test of the high-fidelity CNN model on alloys with different degree of CSRO.** The predictive power—mean absolute error (MAE)—of the high-fidelity CNN model for vacancy migration energy barriers in the equal-atomic TaNbMo alloys with different degree of CSRO. The number of MD steps involved in the hybrid MC/MD simulation to generate the sample with different degree of CSRO is denoted on each panel. Please refer to Supplementary Fig. 4 for the specific degree of CSRO. For test purpose, 800 atoms were randomly chosen from each sample and 8 barriers of each of the 800 atoms were calculated with the NEB method.

To demonstrate that our CNN model is also valid for alloys with different degrees of CSRO, we generated a series of equal-atomic TaNbMo alloys with different degrees of CSRO through hybrid MC/MD simulations, see supplementary Fig. 4. As can be seen from Fig. 4, the test MAE of our high-fidelity CNN model on three samples with much different degrees of CSRO are also impressively low and close to the validation MAE. These confirm that our CNN model is indeed valid for the alloys with different degrees of CSRO. We then used the CNN model to predict the migration energy barrier spectra of a series of states of the equal-atomic TaNbMo alloy during the MC/MD simulation. The evolution of average value, standard deviation and overall VMEB distribution of this concentrated alloy with increasing the degree of CSRO are presented in Supplementary Fig. 5 and Supplementary Fig. 6, which will be useful for optimizing processing condition to tune CSRO for various application of this alloy system. In terms of predicting efficiency, the CNN model has lowered the computational cost by a factor of ~500, as compared with the standard NEB method, see detailed discussion in Supplementary Note 1.

**Compositional dependence of migration barrier spectra**. We have generated $\Delta E_{ij}$ spectra of alloys at 52 different compositions each containing 2,000 atoms in the ternary alloy system to construct the training/validation/test datasets. One would be naturally wondering whether it is possible to extract from the big database some composition-property relationships valuable to alloy design. We indeed got some interesting composition-property relationships in the ternary alloy system through analyzing this database. As seen from Fig. 5, for all three species of moving atoms, the concentration of Mo element plays the dominating role in determining the sample-averaged $\Delta E_{ij}$ and the standard deviation of $\Delta E_{ij}$ is larger around the composition of $Ta_{45}Nb_{10}Mo_{45}$. These composition-property relationships will provide guidance on tuning all properties influenced by diffusion. This highlights the importance of developing such databases for different alloy systems. For this single ternary alloy system modeled by an empirical potential, it is affordable to directly use the NEB calculation to develop this database. However, to build such databases for many different ternary alloy systems, or even quaternary/quinary systems simulated with quantum mechanics-based interatomic fields which are more accurate yet more expensive, it would be impractical to directly use the NEB calculation to construct the entire database and the DL framework presented in this work will be a valuable tool to accelerate the development of these desirable databases, as it is also feasible to train a CNN model using a much smaller training dataset, which would sacrifice a little bit of predictive performance, see Supplementary Note 2 for more discussion.

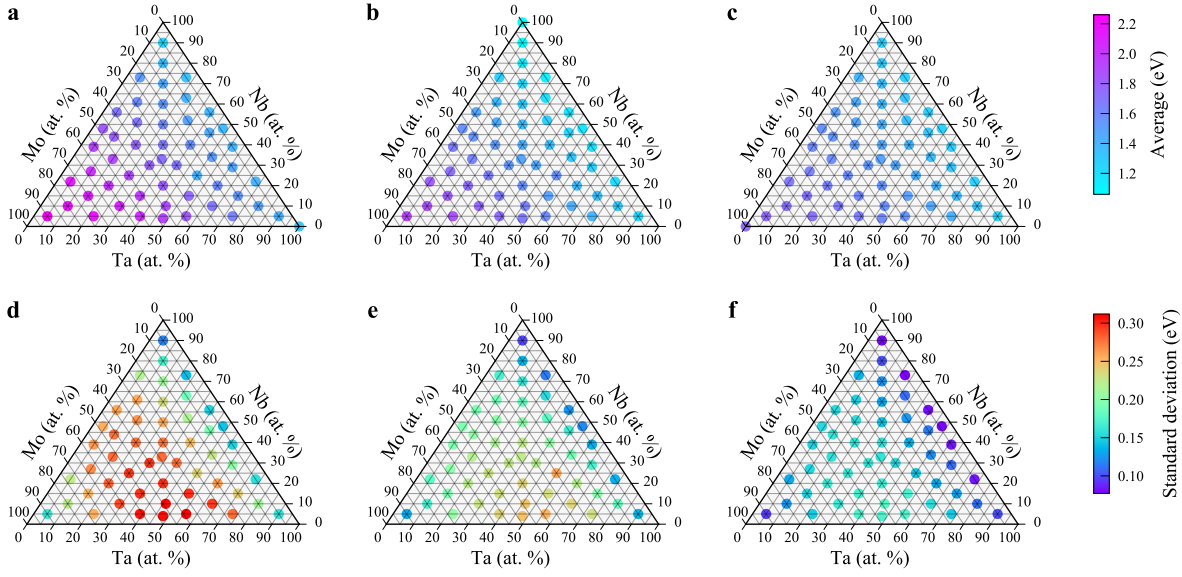

**Fig. 5 Compositional dependence of average and standard deviation of vacancy migration energy barriers. a**, **b** and **c** show the average migration barrier of Ta, Nb and Mo speices, spanning the whole compositional space, respectively. **d**, **e**, and **f** present the corresponding standard deviation of energy barrier spectrum.

In addition to average and standard deviation of $\Delta E_{ij}$, the $\Delta E_{ij}$ distribution can provide more details regarding diffusion behaviors at various compositions. As an example, we explored the influence of Mo concentration on the variation of $\Delta E_{ij}$ distribution of all three species using a series of alloys of Ta$_{(100-x)/2}$Nb$_{(100-x)/2}$Mo$_x$ ($x \in \{34, 35, 36, ..., 99\}$). For these alloys, when the Mo concentration is very high, the number of Ta or Nb will be very low in a small sample containing, say, 2,000 atoms and the resultant $\Delta E_{ij}$ distributions for Ta or Nb species would be incomplete. To avoid this issue, we created large samples for these compositions to ensure that the number of atoms of any species in each sample is not less than 5,000. Then, we used the high-fidelity CNN model to predict the $\Delta E_{ij}$ distributions of these samples (the CNN model has been demonstrated to be reliable to predict $\Delta E_{ij}$ of large samples, see Supplementary Fig. 3). The dependence of both average and standard deviation of $\Delta E_{ij}$ on Mo concentration revealed from these alloys (see supplementary Fig. 8) is in line with that shown in Fig. 5. As seen from supplementary Fig. 9, the $\Delta E_{ij}$ of all species show a bimodal distribution when Mo concentration is very high. And for the Mo species of moving atoms, although sample-averaged $\Delta E_{ij}$ is the highest in pure Mo metal, there are many local regions in both dilute and concentrated alloys which have higher $\Delta E_{ij}$ than that in pure Mo metal. The big data of $\Delta E_{ij}$ spectra empowered by our CNN model will also be valuable to explore the quantitative relationship between energy barrier spectrum and diffusion heterogeneity in HEAs, which may be the focus of our future study.

**Discussion**

The CNN model developed in the present work has been demonstrated to be accurate and efficient in predicting path-dependent $\Delta E_{ij}$ of alloys with different degrees of CSRO over the entire ternary alloy system. And we showcased that the big $\Delta E_{ij}$ database can shed light on the relationships between compositions and properties over a large compositional space. The DL framework proposed in this work will be a valuable tool to develop a spectral diffusion database covering many multi-component alloy systems, which is anticipated to be a general and broadly applicable alloy design and processing optimization toolbox relevant to all material properties impacted by diffusion and thus can accelerate alloy screening for the discovery of desirable properties. The length scale of local chemical environments relevant to VMEB was also uncovered during developing the CNN model, which enhances our understanding of atomic-level structure-property relations and also provides a new methodology to determine the length scale of structure relevant to other properties, e.g., dislocation dynamics. The success of predicting VMEB also has implications to other aspects of materials science. For example, CSRO[35,36] in HEAs is also an intensive research topic. The CSRO in almost all previous atomistic simulations[49,50] were tuned via the swap MC algorithm[51], in which the formation of CSRO is purely determined by thermodynamics and diffusion kinetics is ignored. It is thus reasonable to speculate that the currently formed CSRO in model HEAs may remain different from that in real materials. Our current success of predicting VMEB implies that it is promsing to develop CNN models to instantly predict the barrier of swapping any pair of atoms within a given distance. Thus, the kinetics can be taken into account when implementing DL-enabled kinetic MC algorithm, which will make diffusion-mediated chemical order formation more realistic. The DL framework is also powerful to predict rotation-invariant properties in both crystals and glasses through data augmentation[37], which will be reported elsewhere. The DL framework may also be useful to developing more accurate and efficient machine learned-interatomic potentials[52-54], which deserves further exploration.

## Methods

**Random solid solutions and alloys with chemical short-range orders.** We prepared random solid solution models by randomly assigning atom type (Ta, Nb, Mo) in the system. The portions of atoms being assigned each time depend on targeting concentrations. As this way, we obtained an ensemble of random TaNbMo solid solutions with different concentrations. To prepare a system with CSRO, we use a MC swap of atoms coupled with MD simulations to lower potential energy and increase CSRO. The MC/MD simulation is performed at 300 K with Nose-Hoover thermostat and calls 30 MC trials every 100 MD timesteps to perform trials of exchanging each pair of elements. The atom swaps are accepted or rejected based on the Metropolis algorithm. Sufficient MC/MD steps have been conducted to ensure the convergence of the system potential energy. The non-proportional number is used to quantify the degree of CSRO. The order parameter between any pair of species $\alpha$ and $\beta$ ($\alpha$ represents the species of the central atom and $\beta$ the species of neighboring atoms in the $k$th shell) is defined as $\delta_{\alpha\beta}^k = N_{\alpha\beta}^k - N_{0,\alpha\beta}^k$, where $N_{\alpha\beta}^k$ denotes the actual number of pairs in the $k$th shell, and $N_{0,\alpha\beta}^k$ the number of pairs for the pure random mixture. The values of all pairs in our random system are nearly zero, see supplementary Fig. 4, verifying the random nature of our initial model. The visualization of atomic configurations was realized using the OVITO package[55].

**Vacancy migration energy barrier calculation.** We performed transition state calculations with the nudged elastic band (NEB) method[34] implemented in the LAMMPS package[56] to search the first order saddle point on potential energy surface for vacancy migration to its first nearest neighbor site. Each calculation requires two atomic configurations, i.e., the initial and final configurations. For the initial configuration, we pick an atom $i$ as a center atom and delete it to create a vacancy. We separately move each of the eight first-nearest neighboring atom $j$ to the vacant site, thus creating eight final configurations corresponding to eight migration pathways. For each pathway, we performed energy minimization and box relaxation (to zero pressure) to both initial and final configurations. After that, NEB calculations are performed to obtain the minimum energy path connecting the initial and final states, from which the saddle point and migration energy barrier can be extracted. The NEB spring constant is 5 eV/Å and the stopping tolerance for the energy and force are 0 eV and 0.001 eV/Å, respectively. By looping atom index $i$ and its first-nearest neighbor *j*, the migration barriers at all sites along each of the eight pathways can be obtained.

**Training procedure of CNN models.** We minimized the MAE between true and predicted vacancy migration energy barriers during training the CNN models, and used early stopping and selected the model with the lowest MAE in the validation dataset. The learning rate started from 0.001 and was then divided by $\sqrt{10}$ once the validation MAE plateaued. We chose RMSprop optimizer and mini-batch size of 160, 100, 80 and 60 for CNN models containing 4, 8, 12 and 16 convolutional layers, respectively. The training was implemented in the TensorFlow package[57].

**NN and LR models.** We intensively optimized the architecture of our NN models and found that the optimal model contains two hidden layers and each hidden layer contains 10 neurons activated with the ReLU nonlinearity[37]. And the training procedure of NN models are similar as those for training the CNN models and was also implemented in the TensorFlow package[57]. The training of LR models was implemented in the Sci-kit learn package[58].

**Data availability**
All relevant data are included within the manuscript and Supplementary Information.

**Code availability**
We used open source packages LAMPPS and TensorFlow for our atomistic simulations and deep learning, respectively. All custom codes that support the findings of this study are available from Z. Fan (zfan2016@gmail.com) and B. Xing (bxing2@uci.edu) upon reasonable request.

**Acknowledgements**
This work was supported by the U.S. Department of Energy (DOE), Office of Basic Energy Sciences, under Award DE-SC0022295. B. X. acknowledges support from the National Science Foundation Materials Research Science and Engineering Center program through the UC Irvine Center for Complex and Active Materials (DMR-2011967). The authors also acknowledge the start-up support from the Henry Samueli School of Engineering, University of California, Irvine

**Author contributions**
P.C., Z.F. and B.X. conceived the original research idea. Z.F. constructed the training/validation/test datasets from raw atomistic configurations, carried out the training of machine/deep learning models and analyzed the results. B.X. performed the atomistic simulations and energy barrier calculation. Z.F. led the writing of the first draft of the paper, and P.C. revised the paper with inputs from Z.F. and B.X. All authors contributed to the discussions.

**Competing interests**

The authors declare no competing interests.

**Additional information**
**Supplementary information** The online version contains supplementary material available at xxxxx.

Supplementary Information for

# Convolutional neural networks enable high-fidelity prediction of path-dependent diffusion barrier spectra in multi-principal element alloys

Zhao Fan[1,*], Bin Xing[2], Penghui Cao[1,*]

[1]Department of Mechanical and Aerospace Engineering, University of California, Irvine, Irvine, California 92697, United States
[2]Department of Materials Science and Engineering, University of California Irvine, Irvine, CA 92697, USA
[*] Emails: zfan2016@gmail.com;  caoph@uci.edu

# Supplementary Notes

**Supplementary Note 1: The efficiency of the CNN model in predicting $\Delta E_{ij}$**

It is very efficient to use the CNN model to predict $\Delta E_{ij}$ on samples of arbitrary size at any compositions within the ternary compositional space, since the CNN model treats each local configuration independently and the prediction can be parallelized easily. When serially predicting 4,000 $\Delta E_{ij}$, it currently takes ~7 minutes to convert 4,000 local configurations into SDMs with a single CPU core and less than 30 seconds to predict the 4,000 $\Delta E_{ij}$ from SDMs with a single GPU node. In contrast, it requires about 600 CPU core-hours to calculate the 4,000 $\Delta E_{ij}$ using the NEB method. And we believe that there should be vast space to develop a more efficient programming code to convert local configurations into SDMs and thus further improve the efficiency of predicting $\Delta E_{ij}$ from local configurations using our DL framework, as we have not spent much effort in optimizing this code.

**Supplementary Note 2: Training CNN models on smaller training datasets**

To demonstrate the impressive predictive performance of our DL framework, we used much computational resource to construct a big training dataset and then trained a deep CNN model. However, it is also practical to use a smaller training dataset to train a relatively shallow CNN model when a very low MAE is not required. It can save a lot of computational cost but only sacrifice a little bit of predictive performance. To showcase it, we trained a CNN model containing eight convolutional layers using a much smaller dataset. Specifically, we only selected 218 barriers randomly from each of the 46 compositions to construct a small training dataset of ~10,000 instances. The validation dataset is exactly the same as the one used to develop the high-fidelity CNN model. The lowest validation MAE achieved with this shallow CNN model is 0.0312 eV. And the MAE of this CNN model on the nine test samples are comparable to or even lower than the validation MAE, see Supplementary Fig. 7. These suggest that shallow CNN models trained on small datasets are also acceptable for some cases. Here we simply chose local configurations at random from each sample to construct the small training dataset. One may be able to use a further smaller training dataset to train a CNN model with similar predictive performance as that of our current shallow CNN model using the strategy suggested in ref.[1]. That is, first partitioning all available local configurations into several clusters using some unsupervised machine learning algorithms, such as k-means clustering adopted in ref.[1], and then utilizing the resultant cluster centroids to choose local configurations to construct a further smaller training dataset. This strategy can maximize the difference among local configurations in the training dataset and thus make training more efficient.

# Supplementary Figures

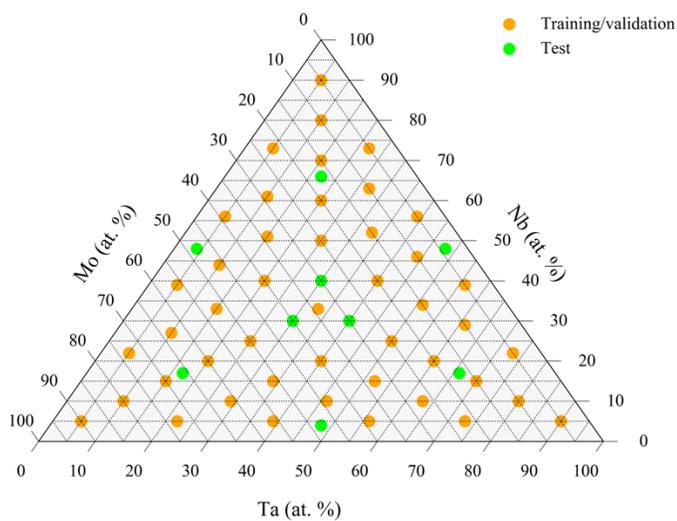

**Supplementary Fig. 1** The specific compositions of the 46 alloys used to construct training/validation datasets are marked with the orange solid circles while the green solid circles denote the nine compositions used to test both the high-fidelity (Fig. 3) and shallow (Supplementary Fig. 7) CNN models.

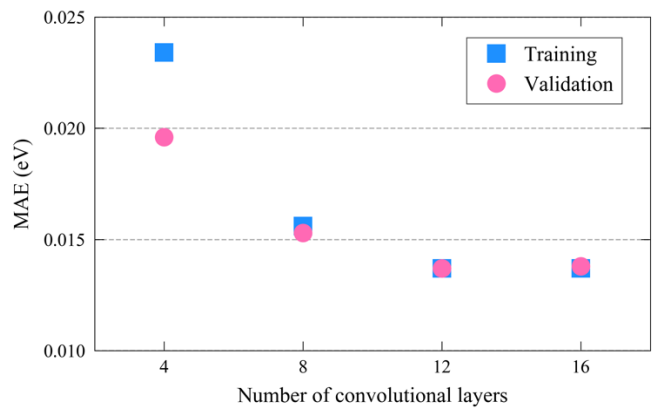

**Supplementary Fig. 2 Elevated predictive power through deepening CNNs.** The mean absolute error (MAE) achieved on the same training/validation dataset constructed using the 46 samples at different compositions via CNN models containing different numbers of convolutional layers.

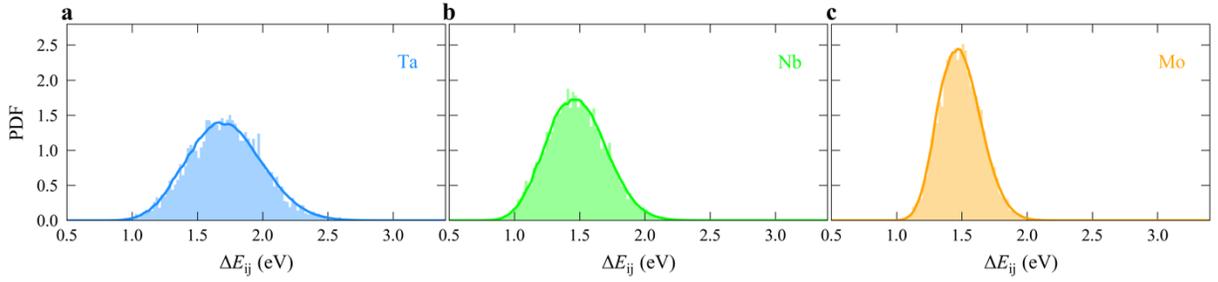

**Supplementary Fig. 3 Comparison between true and predicted distributions of vacancy migration energy barriers.** The bar charts show the true distributions of vacancy migration energy barriers ($\Delta E_{ij}$) for different species of the moving atoms in an equal-atomic TaNbMo random solid solution containing 2,000 atoms. **a**, **b** and **c** correspond to Ta, Nb and Mo species, respectively. The solid curves display the corresponding distributions predicted by the high-fidelity CNN model on a very large equal-atomic TaNbMo random solid solution containing 1,024,000 atoms.

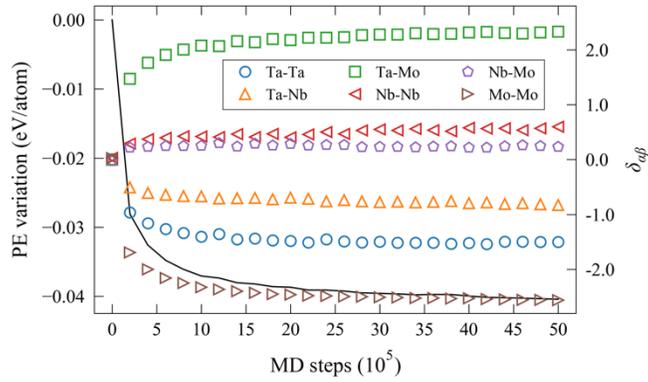

**Supplementary Fig. 4** The evolution of inherent potential energy (PE) (left axis) and degree of chemical short-range order $\delta_{\alpha\beta}$ (see definition in Methods) (right axis) of an equal-atomic TaNbMo alloy starting as a random solid solution during the process of the hybrid MC/MD simulation. The sample contains 16,000 atoms.

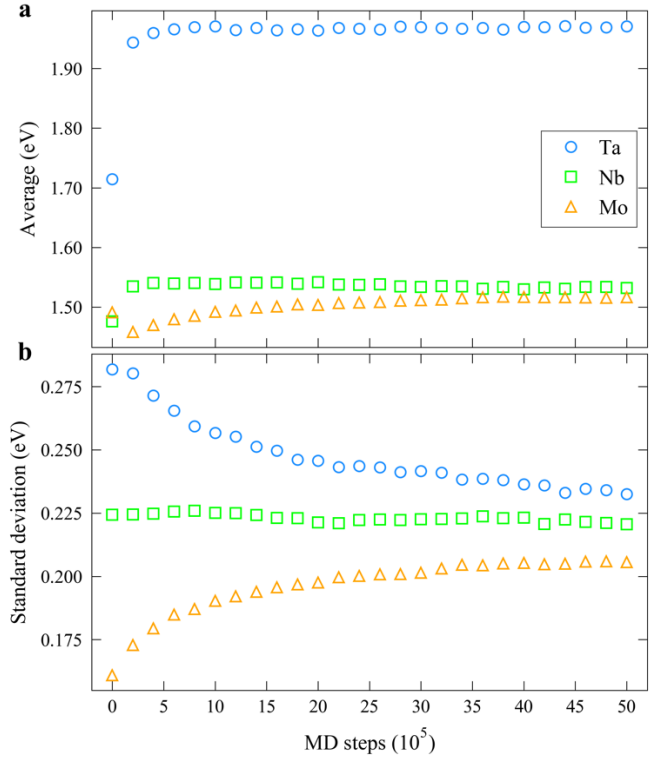

**Supplementary Fig. 5** The average (**a**) and standard deviation (**b**) of vacancy migration energy barriers ($\Delta E_{ij}$) for different species of moving atoms in the equal-atomic TaNbMo alloys with different degree of CSRO predicted by the high-fidelity CNN model as a function of MD steps involved in the hybrid MC/MD simulations. Please refer to Supplementary Fig. 4 for the specific degree of CSRO of these samples.

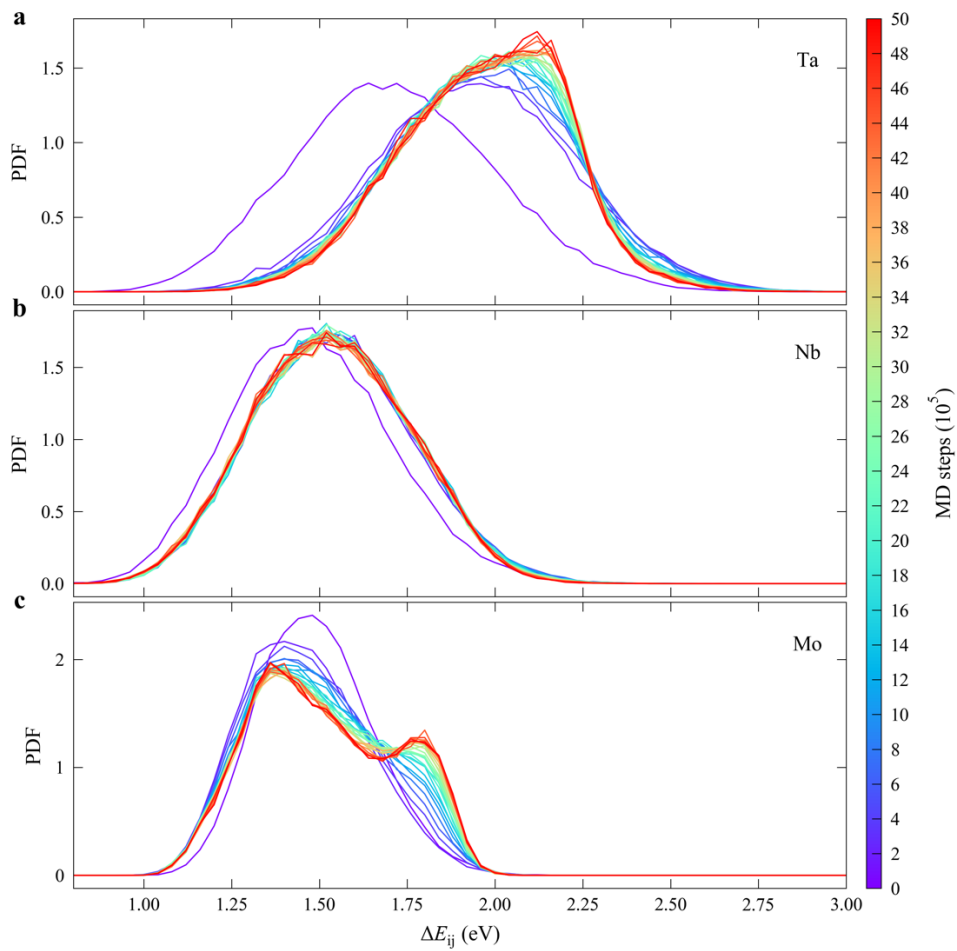

**Supplementary Fig. 6** The distributions of vacancy migration energy barriers ($\Delta E_{ij}$) for different species of moving atoms in the equal-atomic TaNbMo alloys with different degree of CSRO predicted by the high-fidelity CNN model. **a**, **b** and **c** correspond to Ta, Nb and Mo species of moving atoms, respectively. Please refer to Supplementary Fig. 4 for the specific degree of CSRO of these samples.

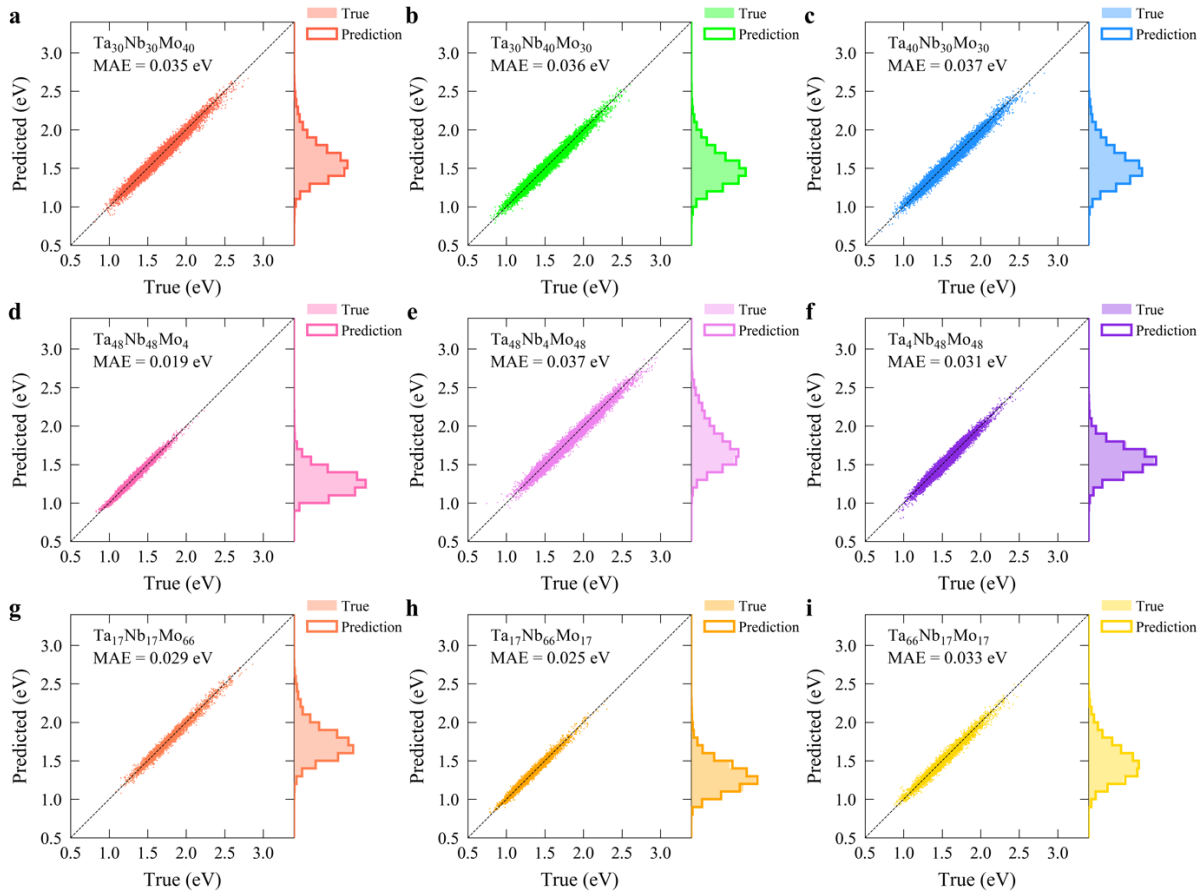

**Supplementary Fig. 7 Test of the shallow CNN model across the compositional space.** The predictive power—mean absolute error (MAE)—of the CNN model trained on the small training dataset containing 10,000 instances for vacancy migration energy barriers in nine alloys at different compositions, which are never used in either training or validation dataset. The specific MAE for each composition is denoted on each panel. Each of the first six samples (**a-f**) contains 2,000 atoms (16,000 barriers) while each of the last three samples (**g-i**) contains 686 atoms (5,488 barriers).

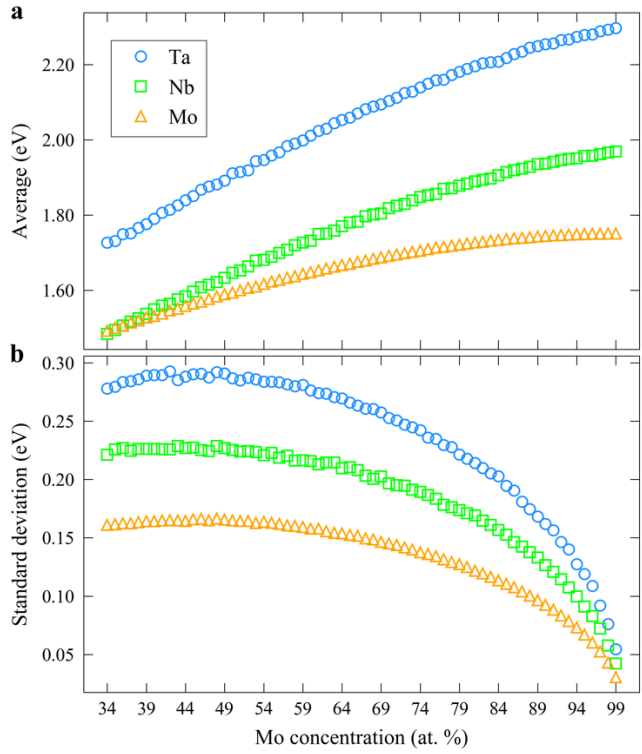

**Supplementary Fig. 8** The average (**a**) and standard deviation (**b**) of vacancy migration energy barriers ($\Delta E_{ij}$) for different species of moving atoms in Ta$_{(100-x)/2}$Nb$_{(100-x)/2}$Mo$_x$ alloys (here $x \in \{34, 35, 36, ..., 99\}$) predicted by the high-fidelity CNN model as a function of Mo concentration.

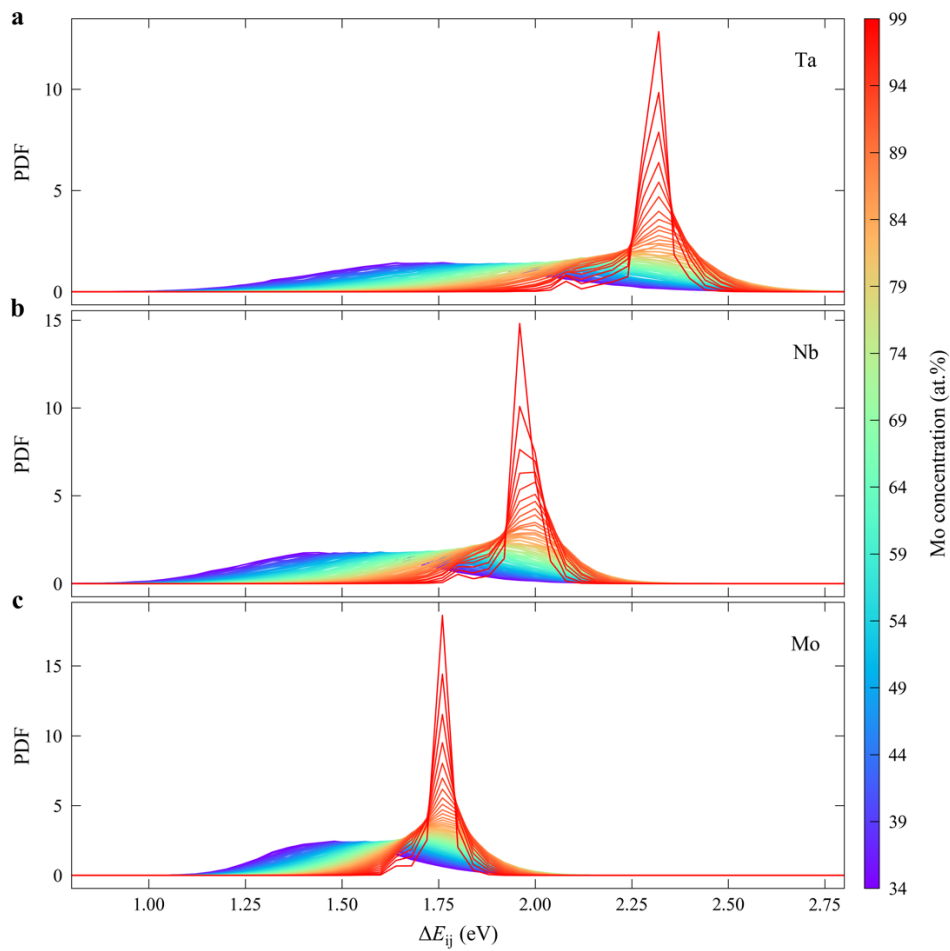

**Supplementary Fig. 9** The distributions of vacancy migration energy barriers ($\Delta E_{ij}$) for different species of moving atoms in Ta$_{(100-x)/2}$Nb$_{(100-x)/2}$Mo$_x$ alloys (here $x \in \{34, 35, 36, ..., 99\}$) predicted by the high-fidelity CNN model. **a**, **b** and **c** correspond to Ta, Nb and Mo species of moving atoms, respectively.

**References**
1. Wagih, M., Larsen, P. M. & Schuh, C. A. Learning grain boundary segregation energy spectra in polycrystals. *Nature Communications* **11**, 6376 (2020).